\documentclass[12pt]{article}
\usepackage{amsfonts,amsmath}

\renewcommand{\theequation}{\thesection\arabic{equation}}

\makeatletter \@addtoreset{equation}{section} \makeatother
\begin{document}

\setlength{\unitlength}{1mm}

\begin{flushright}
FIAN/TD/13/04\\
\end{flushright}

\vspace{0.5cm}

\begin{center}
{\large \bf On Dual Formulation for Higher Spin Gauge Fields in $(A)dS_d$}

\vspace{0.7cm} A.S. Matveev and M.A. Vasiliev

\vspace{0.4cm} \emph{I. E. Tamm Theory Department, Lebedev
Physical Institute,}

\emph{Leninsky prospect 53, 119991, Moscow, Russia}

\vspace{0.4cm}

{\small \texttt{matveev@lpi.ru, vasiliev@lpi.ru}}

\end{center}

\vspace{0.4cm}

\begin{abstract}
\noindent We obtain dual actions for spin $s \geq 2$ massless
fields in $(A)dS_d$ by solving different algebraic constraints in
the same first-order theory. Flat space dual higher spin actions
obtained by Boulanger, Cnockaert and Henneaux~\cite{BH} by solving
differential constraints are shown to result from our formulation
in a sort of quasi-classical approximation for the flat limit. The
case of $s=2$ is considered in detail.
\end{abstract}

\section{Introduction} \label{S:Gen}

The study of dual formulations of classically equivalent field
theories has long history (see e.g.
\cite{OP,CF,DTS,FT,TC,CH,MH,BB,BH} and references therein) and is
important for a deeper understanding of different faces of
superstring theory. In the paper \cite{BH} it was shown that the
dual formulations \cite{TC,CH} of totally symmetric higher spin
(HS) massless fields in flat space can be conveniently obtained
from the first-order formalism developed in \cite{V80,LV}. The
idea was to interpret the equations for the frame-like HS fields
$e$ (just the frame $e_\mu{}^a$ for the spin-2 case) as
differential constraints on the Lorentz connection-type fields
$\omega$. Solving these differential constraints and plugging the
resulting expressions back into the original actions the authors
of \cite{BH} derived dual formulations for the HS fields.

The aim of this paper is to extend the analysis of \cite{BH} to
the $(A)dS_d$ background to show in particular that, for non-zero
background curvature $\lambda^2$, the equations for spin $s \geq
2$ frame-like fields are algebraic constraints $\lambda^2 e +
\partial\omega =0$ allowing $e$ to be expressed in terms of derivatives of
the Lorentz connection-type fields $\omega$. Plugging the
resulting expression back into the original HS action one obtains
equivalent description of the HS dynamics in terms of $\omega$.
The main comment of these notes is that the flat limit description
of \cite{BH} results from our description within the stationary
phase expansion with respect to small $\lambda^2$. The deformation
to $(A)dS_d$ is therefore useful for the analysis of dualities
avoiding procedures of resolving differential restrictions
inevitable in flat space. Our analysis is based on the results of
\cite{LV} where the gauge form of the description of $4d$ HS
dynamics of \cite{V80} was extended to the case of $AdS_d$.

The content of the rest of the paper is as follows. In  Section~2,
by using the MacDowell-Mansouri formulation of gravity with
cosmological term \cite{MM}, we obtain a dual action for a spin-2
field propagating on the $AdS_d$ background and then take the flat
limit within the stationary phase approximation in $\lambda^2$. In
Section~3 dual actions for arbitrary spins $s \geq 2$ in $(A)dS_d$
are obtained and their flat limit is analysed.  Conclusions are
given in Section~4. Appendix contains formulae for
$\tau$-operators of~\cite{LV} used in this paper.

\section{Spin-2 case} \label{S:sp2}

\subsection{MacDowell-Mansouri formulation} \label{S:sp2-1}

As shown by MacDowell and Mansouri~\cite{MM}, $4d$ gravity with
the negative cosmological constant can be formulated in terms of
gauge fields of the $AdS_4$ algebra $o(3,2)$. It was shown
in~\cite{MM} that, up to a topological term, Einstein-Hilbert
action with the cosmological constant admits an equivalent form
\begin{equation}
S^{M} = -\frac{\lambda^{-2}}{4 \kappa^{2}} \int \varepsilon_{abcd}
\,\mathcal{R}^{ab} \wedge \mathcal{R}^{cd} \,,
\end{equation}
with
\begin{equation}
\mathcal{R}^{ab}=d \Omega{}^{ab} + \Omega{}^a{}_c \wedge
\Omega{}^{cb} - \lambda^2 E^a \wedge E^b\,,
\end{equation}
where $E^a$ is the frame 1-form and $\Omega$ is the Lorentz
connection 1-form.

The straightforward $d$-dimensional generalization of the $4d$
MacDowell-Mansouri pure gravity action is~\cite{VGM}
\begin{equation} \label{E:s2-dMM}
S = -\frac{\lambda^{-2}}{4(d-3)!} \int \varepsilon_{c_1 \ldots
c_{d-4} abfg} E^{c_1} \wedge \ldots \wedge E^{c_{d-4}} \wedge
\mathcal{R}^{ab} \wedge \mathcal{R}^{fg}.
\end{equation}
An $AdS_d$ space of radius $\lambda^{-1}$ is described by the
frame $E=h$ and Lorentz connection $\Omega=\omega_0$ satisfying
the equations
\begin{subequations} \label{E:ads-bg}
\begin{align}
d \omega_0{}^{ab} + \omega_0{}^a{}_c \wedge \omega_0{}^{cb} -
\lambda^2 h^a \wedge h^b &=0 , \\
d h^a + \omega_0{}^a{}_c \wedge h^c &=0,
\end{align}
\end{subequations}
where $a, b = 0 \div d - 1$.  Clearly, any solution of
(\ref{E:ads-bg}) is a solution of the field equations of the
action (\ref{E:s2-dMM}). In fact, it is the most symmetric vacuum
solution of the theory~\eqref{E:s2-dMM}. The de Sitter case is
considered analogously with negative $\lambda^2$. For definiteness
we will refer below to the $AdS$ case.

The perturbative expansion near the $AdS$ background is
$$
\Omega=\omega_0+\omega, \quad E=h+e,
$$
where $e$ and $\omega$ are dynamical (fluctuational) parts of the
gauge fields. Then the linear part of $\mathcal{R}^{ab}$ is
\begin{align}
R^{ab} &= D\omega^{ab} - \lambda^2 (h^a \wedge e^b - h^b \wedge
e^a), \\
\intertext{with}%
D\omega^{ab}&=d \omega^{ab} +\omega_0{}^a{}_c \wedge
\omega^{cb}+\omega_0{}^b{}_c \wedge \omega^{ac}, \notag
\end{align}
i.e. $D$ is the background Lorentz covariant differential.

Hence, $d$-dimensional free action of a massless spin-2 field is
\begin{equation} \label{E:s2action}
S^{(2)} = -\frac{\lambda^{-2}}{4(d - 3)!} \int \varepsilon_{c_1
\ldots c_{d-4} abfg} h^{c_1} \wedge \ldots \wedge h^{c_{d-4}}
\wedge R^{ab} \wedge R^{fg}.
\end{equation}
It is invariant under the linearized gauge transformations of the
form
\begin{subequations} \label{E:s2-gauge}
\begin{align}
\delta e^a& = D\varepsilon^a -
h_b \varepsilon^{ab}, \\
\delta \omega^{ab} &= D\varepsilon^{ab} -\lambda^2(h^a
\varepsilon^b- h^b \varepsilon^a),
\end{align}
\end{subequations}
where $\varepsilon^a$ and $\varepsilon^{ab}=-\varepsilon^{ba}$ are
arbitrary gauge parameters.

\subsection{Dual action} \label{S:sp2-2}

After conversion of world indices into Lorentz ones by the
background frame field the spin-2 action reads\footnote{Here
$d\mu$ is the invariant volume-element $det(h_{\mu}{}^c) d^dx$.}
\begin{equation} \label{E:s2actred}
S^{(2)} = -\frac{\lambda^{-2}}{16(d-3)} \int d\mu \,
\delta_{abcd}^{mnpq} R_{mn|}{}^{ab} R_{pq|}{}^{cd},
\end{equation}
where
\begin{equation} \label{E:s2curvred}
R_{cd|}{}^{ab} \equiv h_c{}^{\mu} h_d{}^{\nu} R_{\mu \nu|}{}^{ab}
= D_c \omega_{d|}{}^{ab} - \lambda^2 (\delta_c^a e_d{}^b -
\delta_c^b e_d{}^a) - (c \leftrightarrow d)
\end{equation}
and we use notation
$$\delta_{mnpq}^{abcd}=\frac{1}{(d-4)!}
\varepsilon_{e_1 \ldots e_{d-4} mnpq} \varepsilon^{e_1 \ldots
e_{d-4} abcd}.
$$
The vertical dash in $R_{cd|}{}^{ab}$ and $\omega_{d|}{}^{ab}$
separates converted world indices from the Lorentz ones.

From \eqref{E:s2actred} and \eqref{E:s2curvred} one obtains
\begin{align} \label{E:action-2s}
S^{(2)}[\omega, e] &= -\frac{\lambda^{-2}}{4(d-3)} \int d\mu \, [
\delta_{abfg}^{cdhj} D_c \omega_{d|}{}^{ab} D_h
\omega_{j|}{}^{fg} \notag \\
&- 8\lambda^2 (d-3) (\omega_{c|}{}^{ab} +
\delta_c^a \omega_{d|}{}^{bd} - \delta_c^b
\omega_{d|}{}^{ad})D_a e_b{}^c \notag \\
&+ 4\lambda^4 (d-3)(d-2) (e_a{}^a e_b{}^b - e_a{}^b e_b{}^a) ].
\end{align}
Let us introduce the field $Y_{ab|c}$
\begin{equation} \notag
Y_{ab|c}=\omega_{c|ab}+\eta_{ac}
\omega_{d|b}{}^d-\eta_{bc}\omega_{d|a}{}^d.
\end{equation}
Up to a total derivative, the action~\eqref{E:action-2s} can be
rewritten as
\begin{align} \label{E:action-eY}
S^{(2)}[Y, e] &= \int d\mu \, [ Y_{ab|c} Y^{ac|}{}_b -
\frac{1}{d-2} Y_{ab|}{}^b Y^{ac|}{}_c - 2 D_a Y^{ab|}{}_c e_b{}^c \notag \\
&- \lambda^2(d-2)(e_a{}^a e_b{}^b - e_a{}^b e_b{}^a) ].
\end{align}
Variation of the action~\eqref{E:action-eY} with respect to
$e_a{}^c$ gives
\begin{subequations} \label{E:frameviacon}
\begin{align}
\lambda^{2}e_a{}^b &= -\frac{1}{d-2} \left[ D_c Y^{bc|}{}_a +
\frac{1}{d-1} \delta_a^b D_c Y^{cd|}{}_d \right], \\
\lambda^{2}e_a{}^a &= -\frac{1}{(d-2)(d-1)} D_a Y^{ab|}{}_{b}.
\end{align}
\end{subequations}
Using~\eqref{E:frameviacon} one can get rid of the frame field
in~\eqref{E:action-eY} to obtain the second-order action
\begin{align} \label{E:s2finalaction}
S^{(2)}[Y] = &\int d\mu \, \left[ Y_{ab|c} Y^{ac|}{}_b -
\frac{1}{d-2} Y_{ab|}{}^b Y^{ac|}{}_c - \frac{\lambda^{-2}}{d-2}
D_a Y^{ab|}{}_c D_e Y^{ec|}{}_b\right. \notag \\
&+ \left. \frac{\lambda^{-2}}{(d-2)(d-1)} D_a Y^{ab|}{}_b D_c
Y^{cd|}{}_d \right].
\end{align}

\subsection{Flat theory as a classical limit} \label{S:sp2-3}

The generating functional $\mathcal{Z}$ for the
action~\eqref{E:s2finalaction} is
\begin{equation} \label{E:statphas}
\mathcal{Z} = N\int \mathcal{D}Y \, f[Y] \, e^{i\widetilde{S}[Y]},
\end{equation}
where $N$ is a normalization factor,
$$
f[Y]=e^{i \int d\mu \; ( Y_{ab|c} Y^{ac|}{}_b - \frac{1}{d-2}
Y_{ab|}{}^b Y^{ac|}{}_c)}
$$
and
$$
\widetilde{S}[Y]=\frac{\lambda^{-2}}{d-2} \int d\mu \; ( -D_a
Y^{ab|}{}_c D_e Y^{ec|}{}_b + \frac{1}{d-1} D_a Y^{ab|}{}_b D_c
Y^{cd|}{}_d ).
$$
To evaluate this integral in the flat limit $\lambda \rightarrow
0$ one can apply the stationary phase method. Stationary points of
$\widetilde{S}[Y]$ satisfy the equation
\begin{equation} \label{E:extrfields}
D_b D_a Y^{ad|}{}_c-\delta_c^d D_b D_a Y^{ae|}{}_e - (b
\leftrightarrow c)=0.
\end{equation}
It is readily seen that, up to the terms proportional to
$\lambda^2$ which vanish in the flat limit, \eqref{E:extrfields}
admits a solution of the form
\begin{equation} \label{E:extrimsol}
Y_{ab|}{}^d = D_e Y^e{}_{ab|}{}^d,
\end{equation}
where $Y_{eab|d}=Y_{[eab]|d}$ and $[...]$ mean antisymmetrization.

The key point is that the solutions~\eqref{E:extrimsol} are
dominating because $D_a Y^{ab|}_c=\lambda^2 Y_{ec}{}^{b|e}$ and
therefore $\widetilde{S}[Y] \sim \lambda^2$. Contribution from
other stationary solutions is suppressed when $\lambda \rightarrow
0$ since, because of the boundary terms, $\widetilde{S}[Y] \sim
\lambda^{-2}$.

Thus, in the flat limit the generating functional has the form
\begin{equation} \label{E:flatprop}
\mathcal{Z} \propto \left. \int \mathcal{D}X \; e^{i \int d\mu \;
( Y_{ab|c} Y^{ac|}{}_b - \frac{1}{d-2} Y_{ab|}{}^b Y^{ac|}{}_c)}
\right|_{Y_{ab|}{}^d = \partial_e Y^e{}_{ab|}{}^d}
\end{equation}
that is the flat space action is equivalent to
\begin{equation} \label{E:flataction}
S_{fl}[Y] = \int d^dx (\; \partial_d Y^d{}_{ab|c}
\partial_e Y^{eac|}{}_b - \frac{1}{d-2}
\partial_d Y^d{}_{ab|}{}^b \partial_e Y^{eac|}{}_c) \;,
\end{equation}
which is just the action obtained in~\cite{TC,BH}.

We see that in our approach the dual flat space action results
from some sort of quasi-classical approximation with $\lambda$
being a counterpart of Planck constant in quantum mechanics. From
this interpretation it follows in particular that naive
deformation of the dual flat space formulation to the $AdS$ case
with $\lambda \neq 0$ may be difficult unless all-order
corrections of the quasi-classical expansion are taken into
account, that restores the formulation in terms of the Lorentz
connection as a dynamical variable. Hopefully, such interpretation
may help to answer some of the questions on the structure of
interacting dual theories.

\section{Higher spins} \label{S:arbspin}

\subsection{Frame-like formulation} \label{S:arbsp-1}

As shown in \cite{LV}, a totally symmetric massless field of spin
$s$ in any dimension can be described by a collection of
1-forms\footnote{For simplicity throughout this paper we use
convention of~\cite{V80} that symmetrized groups of indices are
designated by one and the same letter and a number of indices is
given in parentheses, e.g. $a(s-1)$ means $\{a_1 \ldots
a_{s-1}\}$.} $dx^{\mu} \omega_{\mu|}{}^{a(s-1),b(t)}$ ($s-1 \geq t
\geq 0$) which are symmetric in the Lorentz indices $a$ and $b$
separately and satisfy
\begin{subequations} \label{E:prop}
\begin{align}
\omega_{k|}{}^{a(s-1),ab(t-1)} &= 0, \label{E:prop1} \\
\omega_{k|}{}^{a(s-3)c}{}_c{}^{,b(t)} &= 0,
\end{align}
\end{subequations}
where symmetrization over $s$ Lorentz indices $a$ is assumed in
the first relation~\eqref{E:prop1}. From \eqref{E:prop} it follows
that all traces of fiber Lorentz indices are also zero.

Note that the 1-forms $\omega_{a(s-1),b(t)}$
satisfying~\eqref{E:prop} are described by the
traceless Young tableaux
\begin{picture}(20,5)(-4,4)
\linethickness{0.25mm}\scriptsize%
\put(0,0){\line(0,1){6}} \put(0,0){\line(1,0){12}}
\put(3,0){\line(0,1){6}} \put(0,3){\line(1,0){15}}
\put(0,6){\line(1,0){15}} \put(9,0){\line(0,1){6}}
\put(12,0){\line(0,1){6}} \put(15,3){\line(0,1){3}}
\put(3.3,1){\shortstack{. . .}} \put(3.3,4){\shortstack{. . .}}
\put(5.5,-2.5){\shortstack{t}} \put(4.5,6.5){\shortstack{s-1}}
\end{picture}
\vspace{0.6cm}

The linearized curvatures have the structure
\begin{equation} \label{E:gencurv}
R^{a(s-1),b(t)}= D \omega^{a(s-1),b(t)} +
\tau_-(\omega)^{a(s-1),b(t)} - \lambda^2
\tau_+(\omega)^{a(s-1),b(t)},
\end{equation}
with
\begin{align}
&\tau_-(\omega)^{a(s-1),b(t)} = \alpha h_c \wedge
\omega^{a(s-1),b(t)c}, \notag \\
&\tau_+(\omega)^{a(s-1),b(t)} = \beta \Pi(h^b \wedge
\omega^{a(s-1),b(t-1)}), \notag
\end{align}
where $\Pi$ is the projection operator to the irreducible
representation described by the traceless Young tableaux of the
Lorentz algebra $o(d-1,1)$ with $s-1$ and $t$ cells in the first
and the second rows, respectively. $\alpha$ and $\beta$ are some
coefficients which depend on $s,t$ and $d$ and are fixed so that
\begin{equation}
(\tau_-)^2 = 0, \qquad (\tau_+)^2 = 0, \qquad D^2 - \lambda^2
\{\tau_-,\tau_+\} = 0. \notag
\end{equation}
The curvatures~\eqref{E:gencurv} are invariant under the
linearized gauge transformations:
\begin{equation} \label{E:gauge}
\delta \omega^{a(s-1),b(t)}= D \varepsilon^{a(s-1),b(t)} +
\tau_-(\varepsilon)^{a(s-1),b(t)} - \lambda^2
\tau_+(\varepsilon)^{a(s-1),b(t)},
\end{equation}
where $\varepsilon^{a(s-1),b(t)}$ are arbitrary gauge parameters
possessing the symmetry and tracelessness properties analogous to
those of $\omega^{a(s-1),b(t)}$ in~\eqref{E:prop}. Explicit
expressions for $\tau_+$ and $\tau_-$ are given in Appendix.

The quadratic action functional for the massless spin-$s$ field
has the form \cite{LV}
\begin{align} \label{E:genaction}
S^{(s)} = &\lambda^{-2} \int \sum
\limits_{p=0}^{s-2}{\lambda^{-2p} \frac{[(p+1)!]^2}{(d+p-3)!}}
\varepsilon_{c_1 \ldots c_d} h^{c_5} \wedge \ldots \wedge h^{c_d} \notag \\
&\wedge R^{c_1 a(s-2), c_2 b(p)} \wedge
R^{c_3}{}_{a(s-2),}{}^{c_4}{}_{b(p)}.
\end{align}
It is fixed up to an overall factor by the condition that its
variation with respect to the extra fields $\omega^{a(s-1),b(t)}$
with $t \geq 2$ is identically zero
\begin{align} \label{E:extravar}
\frac{\delta S^{(s)}}{\delta \omega^{a(s-1),b(t)}} \equiv 0 &
&\text{for $t \geq 2$.}
\end{align}

\subsection{Dual action for higher spins} \label{S:arbsp-2}

Since the extra fields $\omega^{a(s-1),b(t)}$ with $t \geq 2$
contribute to the action only through surface terms, it is enough
to take into account the terms with the curvatures $R^{a(s-1),b}$
and $R^{a(s-1),b(2)}$ in~\eqref{E:genaction}. Using explicit
expressions for the operators $\tau_\pm$ from~\cite{LV} (see
Appendix) one obtains
\begin{subequations}\label{E:scurv}
\begin{align}
R_{a(s-1),b} &= D\omega_{a(s-1),b} + \lambda^2 [ h_b \wedge
e_{a(s-1)} - h_{a} \wedge e_{a(s-2)b} \notag \\
&+\frac{s-2}{d+s-4} (h^c \wedge e_{cba(s-3)} \eta_{aa} - h^c
\wedge e_{ca(s-2)} \eta_{ab}) ]
\end{align}
and
\begin{align}
R_{a(s-1),}{}^{b(2)} &= \frac{\lambda^2}{\sqrt{d(s-1)(d+s-3)}}
\left[ \frac{(s-2)(d+s-3)}{d-2} h^c \wedge
\omega_{a(s-1),c} \eta^{bb} \right. \notag \\
&- (d+s-3)(s-2) h^{b} \wedge \omega_{a(s-1),}{}^{b} \notag \\
&+ (d+s-3)(s-1) h_{a} \wedge \omega_{a(s-2)}{}^{b,b} \notag \\
&- (s-2)(s-1) h^c \wedge \omega_{ca(s-3)}{}^{b,b} \eta_{aa} \notag \\
&+ (s-3)(s-1) h^c \wedge \omega_{ca(s-2),}{}^{b} \delta_{a}^{b} \notag \\
&+ \frac{(s-2)(s-1)}{d-2} h_c \wedge
\omega^{b(2)}{}_{a(s-3),}{}^c \eta_{aa} \notag \\
&- \left. \frac{(d+2s-6)(s-1)}{d-2} h_c \wedge
\omega_{a(s-2)}{}^{b,c} \delta_{a}^{b} \right],
\end{align}
\end{subequations}
where $\eta_{ab}$ is the flat metric.

From \eqref{E:genaction} one obtains
\begin{align} \label{E:symbact}
S^{(s)} = \frac{1}{2(d-3)(d-2)} \int d\mu \, \{
\mathcal{L}_2[\omega] + \lambda^{-2}\mathcal{L}_1[D\omega, e] \},
\end{align}
with
\begin{subequations} \label{E:lagranzh}
\begin{align}
\mathcal{L}_1[e,D\omega] &\cong \frac{d-2}{4} \delta_{c_1 c_2 c_3
c_4}^{d_1 d_2 d_3 d_4} R_{d_1 d_2|}{}^{c_1 a(s-2),c_2} R_{d_3
d_4|}{}^{c_3}{}_{a(s-2),}{}^{c_4} \; , \\
\mathcal{L}_2[\omega] &\cong \lambda^{-4} \delta_{c_1 c_2 c_3
c_4}^{d_1 d_2 d_3 d_4} R_{d_1 d_2|}{}^{c_1 a(s-2),c_2 b} R_{d_3
d_4|}{}^{c_3}{}_{a(s-2),}{}^{c_4}{}_b,
\end{align}
\end{subequations}
where $\cong$ indicates that extra fields have been neglected.
Substitution of \eqref{E:scurv} into \eqref{E:lagranzh} gives
\begin{subequations}
\begin{align}
&\frac{\mathcal{L}_1}{d-2} \cong \frac{\lambda^4
s^2(d-3)(d-2)}{(s-1)^2}(e_{c|a(s-2)}{}^c e_{d|}{}^{a(s-2)d} -
e_{c|a(s-2)}{}^d e_{d|}{}^{a(s-2)c}) \notag \\
&- \frac{2 \lambda^2 s(d-3)}{s-1}(A_{bc|}{}^{b}{}_{a(s-2),}{}^c
e_{d|}{}^{a(s-2)d} + A_{bc|}{}^{c}{}_{a(s-2),}{}^d
e_{d|}{}^{a(s-2)b} \notag \\
&+ A_{cb|}{}^{d}{}_{a(s-2),}{}^c e_{d|}{}^{a(s-2)b}) + \frac14
\delta_{c_1 c_2 c_3 c_4}^{b_1 b_2 b_3 b_4}A_{b_1b_2|}{}^{c_1
a(s-2),c_2} A_{b_3b_4|}{}^{c_3}{}_{a(s-2),}{}^{c_4} \\
\intertext{and}%
&\mathcal{L}_2 \cong
-\frac{(d+s-3)(d-3)+s^2-s-1}{s-1}\omega_{c|}{}^{a(s-1),c} \,
\omega_{d|}{}_{a(s-1),}{}^d \notag \\
&+ (1+(d+s-3)(d-3)(s-1))\, \omega_{c|}{}^{c}{}^{a(s-2),b} \,
\omega_{d|}{}^d{}_{a(s-2),b} - (c \leftrightarrow d),
\end{align}
\end{subequations}
where $A_{cd|}{}^{a(s-1),b} = D_c \omega_{d|}{}^{a(s-1),b} - D_d
\omega_{c|}{}^{a(s-1),b}$. As expected, $\mathcal{L}_2[\omega]$
vanishes for the spin-2 case.

The gauge transformations~\eqref{E:gauge} for
$\omega_{c|a(s-1),b}$ have the form
\begin{align} \label{E:omegagauge}
\delta \omega_{c|a(s-1),b} &= D_c \varepsilon_{a(s-1),b} +
\varepsilon_{a(s-1),bc} + \eta_{cb} \varepsilon_{a(s-1)} -
\eta_{ca} \varepsilon_{a(s-2)b} \notag \\
&+ \frac{s-2}{d+s-4}\eta_{aa} \varepsilon_{a(s-3)cb}  -
\frac{s-2}{d+s-4} \eta_{ba}\varepsilon_{a(s-2)c},
\end{align}
where $\varepsilon_{a(s-1)},\quad \varepsilon_{a(s-1),b}, \quad
\varepsilon_{a(s-1),b(2)}$ are arbitrary traceless parameters
possessing the same symmetry properties as the respective
connection 1-forms $\omega_{a(s-1),b(t)}$.

The field $\omega_{c|}{}^{a(s-1),b}$ contains the following
irreducible Lorentz traceless components

\begin{picture}(65,10)(-35,-2)
\linethickness{0.25mm}\scriptsize%
\put(-7,3){\line(0,1){3}} \put(-7,3){\line(1,0){3}}
\put(-4,3){\line(0,1){3}} \put(-7,6){\line(1,0){3}}
\put(-3,3.7){\shortstack{$\otimes$}}%
\put(0,0){\line(0,1){6}} \put(0,0){\line(1,0){3}}
\put(3,0){\line(0,1){6}} \put(0,3){\line(1,0){12}}
\put(0,6){\line(1,0){12}} \put(9,3){\line(0,1){3}}
\put(12,3){\line(0,1){3}} \put(3.5,4){\shortstack{. . .}}
\put(4.5,6.5){\shortstack{s-1}}\put(13,3.7){\shortstack{$=$}}%
\put(16,0){\line(0,1){6}} \put(16,0){\line(1,0){3}}
\put(19,0){\line(0,1){6}} \put(16,3){\line(1,0){12}}
\put(16,6){\line(1,0){12}} \put(25,3){\line(0,1){3}}
\put(28,3){\line(0,1){3}} \put(19.5,4){\shortstack{. . .}}
\put(21.5,6.5){\shortstack{s}}\put(29,3.7){\shortstack{$\oplus$}}%
\put(32,0){\line(0,1){6}} \put(32,0){\line(1,0){6}}
\put(35,0){\line(0,1){6}} \put(38,0){\line(0,1){3}}
\put(32,3){\line(1,0){12}} \put(32,6){\line(1,0){12}}
\put(41,3){\line(0,1){3}} \put(44,3){\line(0,1){3}}
\put(35.5,4){\shortstack{. . .}} \put(36.5,6.5){\shortstack{s-1}}
\put(45,3.7){\shortstack{$\oplus$}}%
\put(48,-3){\line(0,1){9}} \put(48,0){\line(1,0){3}}
\put(51,-3){\line(0,1){9}} \put(48,-3){\line(1,0){3}}
\put(48,3){\line(1,0){12}} \put(48,6){\line(1,0){12}}
\put(57,3){\line(0,1){3}} \put(60,3){\line(0,1){3}}
\put(51.5,4){\shortstack{. . .}} \put(52.5,6.5){\shortstack{s-1}}
\put(61,3.7){\shortstack{$\oplus$}}%
\put(64.5,3.5){\shortstack{\normalsize traces .}}
\end{picture}

\noindent As follows from~\eqref{E:omegagauge}, the second tableau
with two cells in the second row is pure gauge.

On the other hand the field $B_{bc|a(s-1)} =
\omega_{b|a(s-1),c}-\omega_{c|a(s-1),b}$ has the following
components

\begin{picture}(46,10)(-45,-4)
\linethickness{0.25mm}\scriptsize%
\put(-7,0){\line(0,1){6}} \put(-7,0){\line(1,0){3}}
\put(-7,3){\line(1,0){3}} \put(-4,0){\line(0,1){6}}
\put(-7,6){\line(1,0){3}} \put(-3,3.7){\shortstack{$\otimes$}}%
\put(0,3){\line(0,1){3}}  \put(3,3){\line(0,1){3}}
\put(0,3){\line(1,0){12}} \put(0,6){\line(1,0){12}}
\put(9,3){\line(0,1){3}} \put(12,3){\line(0,1){3}}
\put(3.5,4){\shortstack{. . .}}
\put(4.5,6.5){\shortstack{s-1}}\put(13,3.7){\shortstack{$=$}}%
\put(16,0){\line(0,1){6}} \put(16,0){\line(1,0){3}}
\put(19,0){\line(0,1){6}} \put(16,3){\line(1,0){12}}
\put(16,6){\line(1,0){12}} \put(25,3){\line(0,1){3}}
\put(28,3){\line(0,1){3}} \put(19.5,4){\shortstack{. . .}}
\put(21.5,6.5){\shortstack{s}}\put(29,3.7){\shortstack{$\oplus$}}%
\put(32,-3){\line(0,1){9}} \put(32,0){\line(1,0){3}}
\put(35,-3){\line(0,1){9}} \put(32,-3){\line(1,0){3}}
\put(32,3){\line(1,0){12}} \put(32,6){\line(1,0){12}}
\put(41,3){\line(0,1){3}} \put(44,3){\line(0,1){3}}
\put(35.5,4){\shortstack{. . .}} \put(36.5,6.5){\shortstack{s-1}}
\put(45,3.7){\shortstack{$\oplus$}}%
\put(48.5,3.5){\shortstack{\normalsize traces .}}
\end{picture}

\noindent Comparing the Lorentz tensor patterns of the fields
$B_{bc|a(s-1)}$ and $\omega_{b|a(s-1),c}$ and taking into account
the gauge invariance, one concludes that the
action~\eqref{E:symbact} can be expressed in terms of
$B_{bc|a(s-1)}$. The final result is
\begin{align} \label{E:acteB}
S^{(s)}[e,B] &= \frac{s}{2(d-2)} \int d\mu \, \{ \frac{\lambda^2
s(d-2)^2}{(s-1)^2}(e_{c|a(s-2)}{}^c
e_{d|}{}^{a(s-2)d} - e_{a|a(s-2)}{}^d e_{d|}{}^{a(s-1)}) \notag \\
&+ \frac{4(d-2)}{s-1} D_{[c}e_{d]|}{}^{a(s-1)}
(B_{a}{}^{c|}{}_{a(s-2)}{}^d -\frac{1}{2(s-1)}B^{cd|}{}_{a(s-1)}  \notag \\
&+ 2\delta_a^c B_{e}{}^{d|}{}^{e}{}_{a(s-2)} + (s-2)\delta_a^c
B_{ea|}{}^{e}{}_{a(s-3)}{}^d) \notag \\
&+ (d+s-4)(B_{cb|}{}^c{}_{a(s-2)}B_{d}{}^{b|da(s-2)} +
(s-1)B_{ca|}{}^c{}_{a(s-2)}B_{d}{}^{a|da(s-2)} \notag \\
&+B_{bc|a(s-1)}B^{ab|ca(s-2)} -
\frac{1}{2(s-1)}B_{bc|a(s-1)}B^{bc|a(s-1)}) \}.
\end{align}
Let us introduce the field $Y_{bc|}{}^{a(s-1)}$
\begin{align}
Y_{cd|}{}^{a(s-1)}&=B^{a}{}_{c|d}{}^{a(s-2)}-
\frac{1}{2(s-1)}B_{cd|}{}^{a(s-1)} -
2\delta_{c}^{a}B_{db|}{}^{a(s-2)b} \notag \\
&+(s-2)\delta_{c}^{a}B_{b}{}^{a|}{}_{d}{}^{a(s-3)b} - (c
\leftrightarrow d), \notag
\end{align}
which has the properties
\begin{equation}\label{E:Y-prop}
Y_{cd|}{}^{a(s-1)}=Y_{[cd]|}{}^{a(s-1)} \quad \text{and} \quad
Y_{cd|}{}^{a(s-3)b}{}_b=0.
\end{equation}
Now the action~\eqref{E:acteB} can be rewritten in the form
\begin{align} \label{E:acteY}
S^{(s)}&[e,Y] = \int d\mu \, \left( \frac{\lambda^2
s^2(d-2)}{2(s-1)^2}(e_{c|a(s-2)}{}^c
e_{d|}{}^{a(s-2)d} - e_{a|a(s-2)}{}^d e_{d|}{}^{a(s-1)})\right.\notag \\
&+ \frac{s}{s-1}Y^{cd|}{}_{a(s-1)} D_{[c}e_{d]|}{}^{a(s-1)}
+ \frac{s-1}{2}[-Y_{bc|a(s-1)}Y^{ba|ca(s-2)} \notag \\
&+ \frac{1}{d+s-4}\{(s-3)Y_{bc|a(s-2)}{}^bY^{cd|a(s-2)}{}_d -
(s-2)Y_{bc|a(s-2)}{}^bY^{ad|ca(s-3)}{}_d \}\notag \\
&+ \left.\frac{s-2}{2(s-1)}Y_{bc|a(s-1)}Y^{bc|a(s-1)}]\right).
\end{align}
Variation of the action~\eqref{E:acteY} with respect to
$e_{c|a(s-1)}$ yields
\begin{subequations}
\begin{align}
\lambda^2 e_{a|a(s-2)}{}^d &= -\frac{s-1}{s(d-2)}\left(
D_cY^{cd|}{}_{a(s-1)} - \frac{s-1}{d+s-3}\delta_{a}^d
D_cY^{cb|}{}_{a(s-2)b}\right), \\
\lambda^2 e_{c|}{}^c{}_{a(s-2)} &= \frac{(s-1)^2}{s(d-2)(d+s-3)}
D_c Y^{cd|}{}_{a(s-2)d}.
\end{align}
\end{subequations}
Taking into account that $e_{b|a(s-1)} = s\, e_{(b|a(s-1))} -
(s-1)\, e_{a|a(s-2)b}$ one obtains
\begin{align}
\lambda^2 e_{d|a(s-1)} &= -\frac{s-1}{s(d-2)}\left[
\frac{(s-1)(s-3)}{d+s-3}\delta_{a}^d
D_cY^{cb|}{}_{a(s-2)b} \right. \notag \\
&+ (s-1)D_cY^c{}_{a|a(s-2)d} - (s-2)D_cY^c{}_{d|a(s-1)} \notag \\
&- \left.
\frac{(s-1)(s-2)}{d+s-3}\eta_{aa}D_cY^{cb|}{}_{a(s-3)bd}\right].
\end{align}
Plugging this expression back into~\eqref{E:acteY} one gets the
dual action in terms of the field $Y$
\begin{align} \label{E:finact}
S^{(s)}&[Y] = \frac{1}{2(d-2)} \int d\mu \, \left(
\lambda^{-2}[\frac{(s-1)(s-3)}{d+s-3}
D_cY^{cb|}{}_{a(s-2)b}D_fY^{fd|a(s-2)}{}_d \right. \notag \\
&- (s-2)D_cY^c{}_{d|a(s-1)}D_fY^{fd|a(s-1)} +
(s-1)D_cY^c{}_{e|a(s-1)}D_fY^{fa|a(s-2)e}] \notag \\
&+ (s-1)(d-2)\{-Y_{bc|a(s-1)}Y^{ba|ca(s-2)} +
\frac{s-2}{2(s-1)}Y_{bc|a(s-1)}Y^{bc|a(s-1)}\notag \\
&+ \left. \frac{1}{d+s-4}[(s-3)Y_{bc|a(s-2)}{}^bY^{cd|a(s-2)}{}_d
-(s-2)Y_{bc|a(s-2)}{}^bY^{ad|ca(s-3)}{}_d ]\}\right).
\end{align}
For $s=2$ one recovers the action~\eqref{E:s2finalaction} of
Section~\ref{S:sp2-2}.

\subsection{Flat limit}\label{S:arbsp-3}

The flat limit of the action~\eqref{E:finact} is obtained
analogously to the spin-2 case. Firstly, one writes the equation
for stationary points of the part of the action singular in
$\lambda$:
\begin{align}\label{E:extrim}
0 &=(s-1)D_fD_cY^c{}_{a|a(s-2)d} - (s-2)D_fD_cY^c{}_{d|a(s-1)} \notag \\
&+ \frac{s-1}{d+s-3}[(s-3)D_fD_cY^{cb|}{}_{ba(s-2)}\eta_{ad}
-(s-2)D_fD_cY^{cb|}{}_{bda(s-3)}\eta_{aa}] \notag \\
&- (d \leftrightarrow f) + O(\lambda^2),
\end{align}
where the properties~\eqref{E:Y-prop} were taken into account.

One observes that, up to the terms proportional to $\lambda^2$,
the equation~\eqref{E:extrim} has a solution
\begin{equation}\label{E:extrcond}
Y_{bc|a(s-1)}=D_d Y^d{}_{bc|a(s-1)},
\end{equation}
where
$$
Y_{bcd|a(s-1)}=Y_{[bcd]|a(s-1)}, \quad Y_{bcd|a(s-3)e}{}^e=0.
$$

The solutions~\eqref{E:extrcond} are dominating because $D_b
Y^b{}_{c|}{}^{a(s-1)} \sim \lambda^2$, i.e. no boundary terms
contribute for this case. Hence, the part of the
action~\eqref{E:finact} singular in $\lambda$ becomes proportional
to $\lambda^2$ being restricted to~\eqref{E:extrcond}.
Contribution from other stationary solutions is suppressed due to
non-zero boundary terms.

Thus, up to an overall factor, the action~\eqref{E:finact} is
\begin{align}
S_{fl}[Y] &= \int d^dx \, \{
-\partial_eY^e{}_{bc|a(s-1)}\partial_fY^{fba|ca(s-2)}
+ \frac{s-2}{2(s-1)}\partial_eY^e{}_{bc|a(s-1)}
\partial_fY^{fbc|a(s-1)}\notag \\
&+\frac{1}{d+s-4}[(s-3)\partial_eY^e{}_{bc|a(s-2)}{}^b
\partial_fY^{fcd|a(s-2)}{}_d
\notag \\
&-(s-2)\partial_eY^e{}_{bc|a(s-2)}{}^b
\partial_fY^{fad|ca(s-3)}{}_d]\}.
\end{align}
This is precisely the action of~\cite{BH}.

\section{Conclusion} \label{S:conc}

The main conclusion of this paper is that flat space dual actions
for massless fields of spins $s \geq 2$ result from a kind of
quasi-classical approximation in which inverse $(A)dS_d$ radius
$\lambda$ is interpreted as a small parameter analogous to the
Planck constant in quantum mechanics. For $\lambda \neq 0$, the
action we start with is manifestly equivalent to the original HS
action because it results from resolution of algebraic constraints
which express frame-like fields in terms of (derivatives of)
Lorentz connection-like fields. The resulting dual action contains
inverse powers of $\lambda$, so that its flat limit can be
conveniently analysed within stationary phase approximation. Since
the free dual HS actions are shown to result from the
quasi-classical expansion in $\lambda^2$, our approach can also be
useful for the analysis of dual formulations of interacting
theories in the flat and $(A)dS_d$ cases.

It should be mentioned that, unfortunately, the important case of
the spin-1 field cannot be handled in our approach since there is
no auxiliary Lorentz-like connection for spin-1 (a related
property is that the spin-1 field equations do not contain
curvature-dependent mass-like terms in the $(A)dS_d$ space-time).

\section*{Acknowledgements} \label{S:thanks}

\noindent The authors wish to thank O.V. Shaynkman and all
participants of the informal higher spin seminar for a stimulating
discussion. This research was supported in part by grants INTAS
No.00-01-254, the RFBR No.02-02-17067 and the LSS No.1578.2003-2.
The A.M. work was partially supported by the grant of Dynasty
Foundation and ICFP.

\renewcommand{\theequation}{\arabic{equation}}
\section*{Appendix. Operators $\tau_{\pm}$} \label{S:appendix}

Here we reproduce some relevant formulae of~\cite{LV}.

The fields $dx^{\mu}\omega_{\mu|a(s),b(t)}$ can be conveniently
described in terms of ``generating functions", which are ``state
vectors" in the Fock space generated by the operators $a_n^a$,
$a_m^{+a}$ ($n, m = 1,2$) obeying Heisenberg commutation
relations,
\begin{equation}
[a_n^{a},a_m^{+b}]=\delta_{nm}\eta^{ab}, \qquad
[a_n^{a},a_m^{b}]=[a_n^{+a},a_m^{+b}]=0.
\end{equation}
Given $p$-form $A_{a(s),b(t)}$, we introduce a ``Fock vector"
\begin{equation}
|A(p,s,t)\rangle=\underbrace{a_1^{+a} \ldots a_1^{+a}}_{s}
\underbrace{a_2^{+b} \ldots a_2^{+b}}_{t} A_{a(s),b(t)}|0\rangle.
\end{equation}
Consider the subspace extracted by the conditions
\begin{subequations}
\begin{align}
&a_2^{b}a_{1b}^+|A(p,s,t)\rangle=0, \\
&a_n^{b}a_{mb}|A(p,s,t)\rangle=0, \\
&N_1|A(p,s,t)\rangle=(s-1)|A(p,s,t)\rangle,
\end{align}
\end{subequations}
where $N_m=a_m^{+b}a_{mb}$.

The linearized HS curvatures $|R(2,s,t)\rangle$ are
\begin{equation}\label{E:app-cur}
|R(2,s,t)\rangle=D|\omega(1,s,t)\rangle + \tau_-
|\omega(1,s,t+1)\rangle - \lambda^2 \tau_+
|\omega(1,s,t-1)\rangle.
\end{equation}
Here $D$ is the Lorentz covariant differential with respect to
background Lorentz connection and $\tau_{\pm}$ have the following
form
\begin{subequations}
\begin{equation}
\tau_-=\left[\frac{(d+N_1+N_2-3)(N_1-N_2+1)}{d+2N_2-2}\right]^{1/2}h^b
a_{2b},
\end{equation}
\begin{align}
\tau_+=&[(d+N_1+N_2-4)(N_1-N_2+2)(d+2N_2-4)]^{-1/2} \notag \\
&\times\{(d+N_1+N_2-4)[(N_1-N_2+1)h_b a_2^{+b}-h_b
a_1^{+b}(a_2^{+c}a_{1c})] \notag \\
&+ (a_1^{+b}a_{1b}^+)(a_2^{+c}a_{1c})h^aa_{1a} -
\frac{1}{d+2N_2-6}(a_1^{+b}a_{1b}^+)(a_2^{+c}a_{1c})^2h^aa_{2a} \notag \\
&- (N_1-N_2)(a_1^{+b}a_{2b}^+)h^aa_{1a} +
\frac{d+2N_1-4}{d+2N_2-6}(a_1^{+b}a_{2b}^+)
(a_2^{+c}a_{1c})h^aa_{2a} \notag \\
&-\frac{(N_1-N_2+1)(d+N_1+N_2-4)}{d+2N_2-6}
(a_2^{+b}a_{2b}^+)h^aa_{2a}
\}.
\end{align}
\end{subequations}
All non-polynomial functions of $N_{1,2}$ are understood as usual
by
$$
f(N_1,N_2)|A(p,s,t)\rangle=f(s,t)|A(p,s,t)\rangle.
$$
The curvatures~\eqref{E:app-cur} are invariant under the
linearized gauge transformations \eqref{E:gauge}
\begin{equation}
\delta|\omega(1,s,t)\rangle = D|\varepsilon(0,s,t)\rangle + \tau_-
|\varepsilon(0,s,t+1)\rangle - \lambda^2 \tau_+
|\varepsilon(0,s,t-1)\rangle,
\end{equation}
where $|\varepsilon(0,s,t)\rangle$ is an arbitrary gauge
parameter.


\vspace{1cm}
\begin{thebibliography}{6}

\bibitem{BH}
N. Boulanger, S. Cnockaert and M. Henneaux, \emph{JHEP}
\textbf{0306} (2003) 60, hep-th/0306023

\bibitem{OP}
V.I. Ogievetsky and I.V. Palubarinov, \emph{Iad. Fiz.} \textbf{4}
(1966) 216; \emph{Sov. J. Nucl. Phys.} \textbf{4} (1967) 156

\bibitem{CF}
T. Curtright and P.G.O. Freund, \emph{Nucl. Phys.} \textbf{B172}
(1980) 413

\bibitem{DTS}
S. Deser, P.K. Townsend and W. Siegel, \emph{Nucl. Phys.}
\textbf{B184} (1981) 333

\bibitem{FT}
E.S. Fradkin and A.A. Tseytlin, \emph{Annals Phys.} \textbf{162}
(1985) 31.

\bibitem{TC}
T. Curtright, \emph{Phys. Lett.} \textbf{B165} (1985) 304

\bibitem{CH}
C.M. Hull, \emph{JHEP} \textbf{0109} (2001) 027, hep-th/0107149

\bibitem{MH}
P. de Medeiros and C. Hull, \emph{Commun. Math. Phys.}
\textbf{235} (2003) 255, hep-th/0208155

\bibitem{BB}
X. Bekaert and N. Boulanger, \emph{Phys. Lett. B} \textbf{561}
(2003) 183, hep-th/0301243

\bibitem{V80}
M.A. Vasiliev, \emph{Sov. J. Nucl. Phys.} \textbf{32} (1980) 855;
\emph{Fortschr. Phys.} \textbf{35} (1987) 741

\bibitem{LV}
V.E. Lopatin and M.A. Vasiliev, \emph{Mod. Phys. Lett} \textbf{A3}
(1988) 257

\bibitem{MM}
S.W. MacDowell and F. Mansouri, \emph{Phys. Rev. Lett.}
\textbf{38} (1977) 739

\bibitem{VGM}
M.A. Vasiliev, \emph{Nucl. Phys.} \textbf{B616} (2001) 106,
hep-th/0106200

\end{thebibliography}
\end{document}